\documentclass[9pt]{book}

\usepackage[dvips]{graphicx,color}
\usepackage{makeidx,universe}

\makeauthorindex
\makeindex

\BookTitle{The proceedings of the Physics of Accreting Compact Binaries}

\CopyRight{\copyright 2010 by Universal Academy Press, Inc.}

\begin{document} 
\pagenumbering{arabic}

\chapter{%
CYCLOPS-X: Simultaneous optical and X-ray modeling of polars}

\author{\raggedright \baselineskip=10pt%
{\bf Karleyne M. G. Silva$^{1}$, Claudia V. Rodrigues$^{1}$ and Joaquim E. R. Costa$^{1}$}\\ 
{\small \it %
(1) Divis\~ao de Astrof\'\i sica, Instituto Nacional de Pesquisas Espaciais, Av. dos Astronautas 1758, 12227-010, S\~ao Jos\'e dos Campos-SP Brazil \\
}
}

\AuthorContents{Karleyne M. G. da Silva, Claudia V. Rodrigues and Joaquim E. R. Costa} 

\AuthorIndex{Silva}{K. M. G.} 

\AuthorIndex{Rodrigues}{C. V.} 

\AuthorIndex{Costa}{ J. E. R.} 

     \baselineskip=10pt
     \parindent=10pt

\section*{Abstract} 

From the optical to the X-ray frequencies, most of the continuum emission from AM Her systems originates in the post-shock region. Hence, using multi-spectral data can be useful to restrict the physical and geometrical properties of that region. In spite of that, distinct codes are used to model these frequency ranges. {\sc cyclops} is a code to model the optical cyclotron emission of polars.  In this contribution, we present the first version of {\sc cyclops-x}, an improvement of {\sc cyclops} to fit simultaneously optical and X-ray data. The new code adds the bremsstrahlung as an emission process as well as the effects of X-ray absorption by the upper portion of the accretion column. As a first application of  {\sc cyclops-x}, we present X-ray light curves using two sets of parameters provided by the optical modeling of CP Tuc. These two cases have very similar optical emission, but quite distinct light-curves in high energies. It illustrates the need of simultaneous modeling of the optical and X-ray emission to a proper description of the magnetic accretion in polars. 
 
\section{Introduction} 

Polars are a subclass of the cataclysmic variables (CVs) in which the accreted material is channeled by the magnetic field of the white dwarf (WD) forming an accretion column. Near the WD surface the material forms a shock front increasing temperature and densities. The bulk of the emission comes from the base of the column also called post-shock region. The mean emission processes are cyclotron (optical and infrared) and bremsstrahlung (X-ray). The radiative cooling results in a structured region with temperature and density varying as a function of the height of the post-shock region.

The X-ray emission of polars is commonly reproduced combining shock models and plasma emission models, as Meka model \cite{mewe:1986} available in the {\sc xspec} Spectral Fitting Package \cite{arnoudi:1996}. First, the post-shock region is divided in layers and each layer has a temperature and density given by the post-shock structure. Then, the emissivity is calculated for each layer and the total emission is the sum of the plasma emissivity of all layers, assuming an optically thin region \cite{cropper:2000}.  

The X-ray fittings are usually done without considering the optical data, in part because distinct codes are used to model each spectral range. An exception is the modeling of polar RX J2115-5840 \cite{ramsay:2000}. The cyclotron 2D emitting region, obtained from the fit of optical data, is used to define the location of the X-ray emission. However, the density and temperature profiles adopted in fitting each energy band are not the same. 

The absorption due the pre-shock flow located above the post-shock region is important in high energies. Customary, the absorption, which varies with orbital phase, is arbitrarily choosen in order to fit the observational data. 

The {\sc cyclops} code solves the radiative transfer of the cyclotron process in the post-shock region adopting a 3D treatment and specific functions to describe the variation of temperature and density due the radiative cooling. The code provides the four Stokes parameters which can be used to reproduce optical light and polarimetric curves from polars \cite{costa:2009}. Here we present the modification of {\sc cyclops} to include bremsstrahlung emission and to prepare the code to simultaneously model the optical and X-ray continuum emission of post-shock regions of polars.

\section{CYCLOPS-X}

{\sc cyclops-x} has exactly the same input parameters of {\sc cyclops} \cite{costa:2009}. Hence, no additional parameters are needed to model X-ray data. The post-shock and pre-shock regions are defined by the magnetic field lines and divided in several elements of volume according to the chosen spatial resolution. The emergent flux in a given orbital phase is the sum of the fluxes coming from all lines of sight. The code solves the radiative transfer to each line of sight in steps, considering an inhomogeneous post-shock region. This naturally incorporates the shock structure and optical depth effects.

The emissivity of the bremsstrahlung is calculated in each voxel according to \cite{gronenschild:1978}:
\begin{equation}
\label{emissividade}
j_{\lambda} (N_e, T , \lambda, g) = 2.051 \times 10^{-19} e^{\frac{1.439 \times 10^8}{\lambda T}} g  N_e^{2} \lambda^{-2}  T ^{-0.5},
\end{equation}
where $N_e$ ($cm^{-3}$) is the electronic number density, $\lambda$ (\AA) is the wavelength, T(K) is the temperature and $\textit{g}$ is the non-relativistic Gaunt factor \cite{mewe:1986}. 

The absorption is considered only in the phases in which the pre-shock region is between the observer and the post-shock region. This is done for each line of sight that composes the total flux in a given orbital phase. The extinction processes considered are photoelectric absorption \cite{morrison:1983} ($\sigma_{ph}$) and Thomson scattering ($\sigma_{th}$). The flux observed in each orbital phase ($F_{pha}$) is sum of the emergent flux in a line of sight ($F_0$)  attenuated by the upper part of the column according to:
\begin{equation}
F_{pha}= \sum_{i=0}^{l} F_0^{l} \times e^{-N_e^{l} \ s \ (\sigma_{th}+ \sigma_{ph})},
\end{equation}
where $s$ is the optical path in the line of sight and $l$ is the number of lines of sight used to represent the region. 

\section{An example of CYCLOPS X-ray light curves: CP Tuc}

In order to illustrate the light curves produced by {\sc cyclops-x}, we chose the polar CP Tuc.

Initially, it was suggested that the post-shock region of CP Tuc is always visible and that the modulation observed in the X-ray light curves would be due absorption by the upper accretion column in specific orbital phases \cite{mizaki:1996}. After that, the modeling of the optical light and polarization curves indicated a different geometry to CP Tuc, in which its post-shock region is self-eclipsed (by the WD) during half of the orbital period generating the modulation \cite{ramsay:1999}. Recently, our group presented new data in two optical bands. {\sc cyclops} modeling of these data shows that it is possible to find good models in the two cases above: the absorption and the self-eclipse scenarios. (Please see \textit{Cyclotron modeling of the polar CP Tuc}, Rodrigues et al., this meeting, for more details.)

\begin{figure}[t]
 \begin{tabular}{cc}
  \begin{minipage}{.45\hsize}
   \begin{center}
     \includegraphics[trim=0cm 1.0cm 0cm 1.0cm,clip=true,width=.70\textwidth]{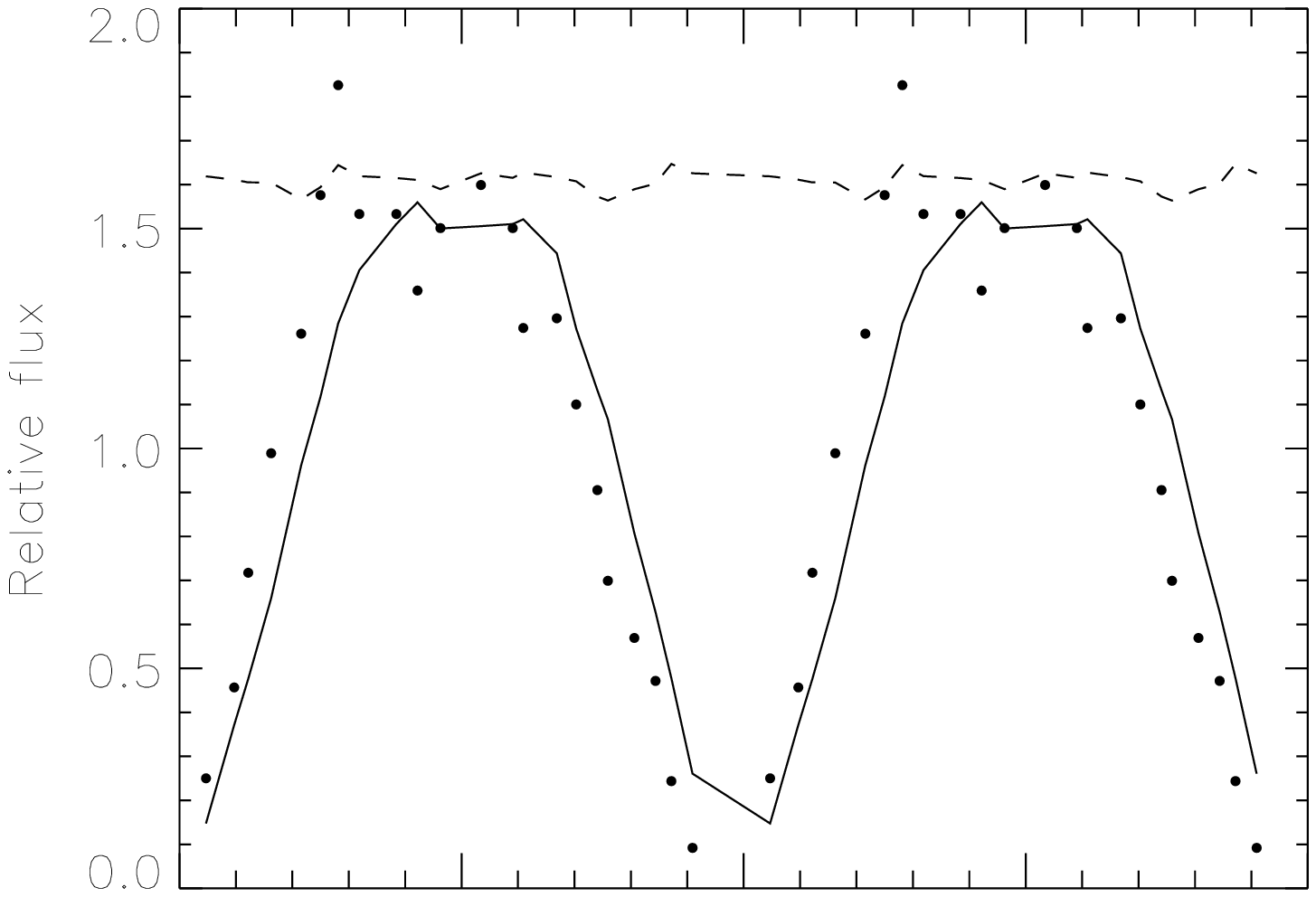}
     \includegraphics[trim=0cm 1.0cm 0cm 1.0cm,clip=true,width=.70\textwidth]{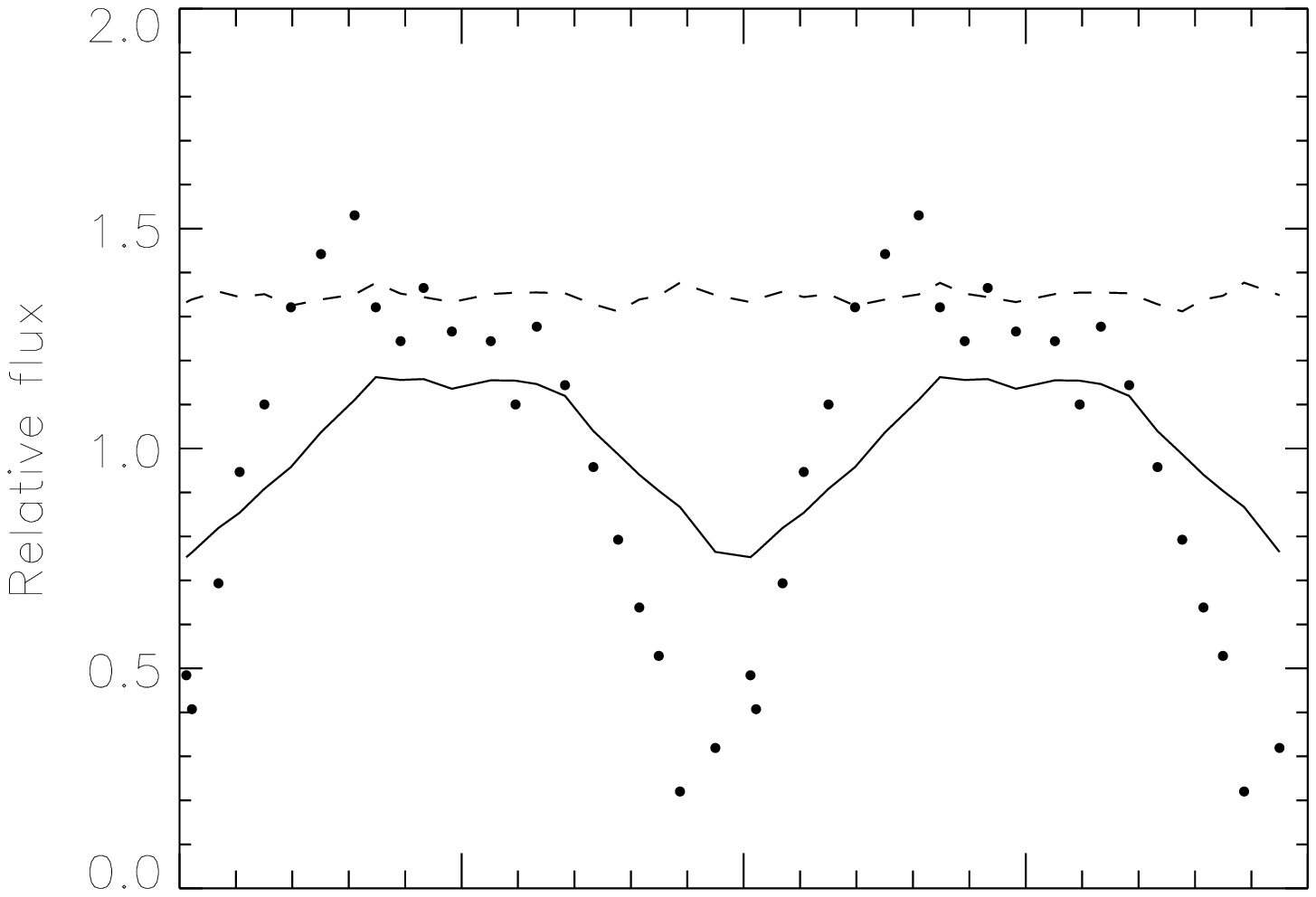}
     \includegraphics[trim=0cm 1.0cm 0cm 1.0cm,clip=true,width=.70\textwidth]{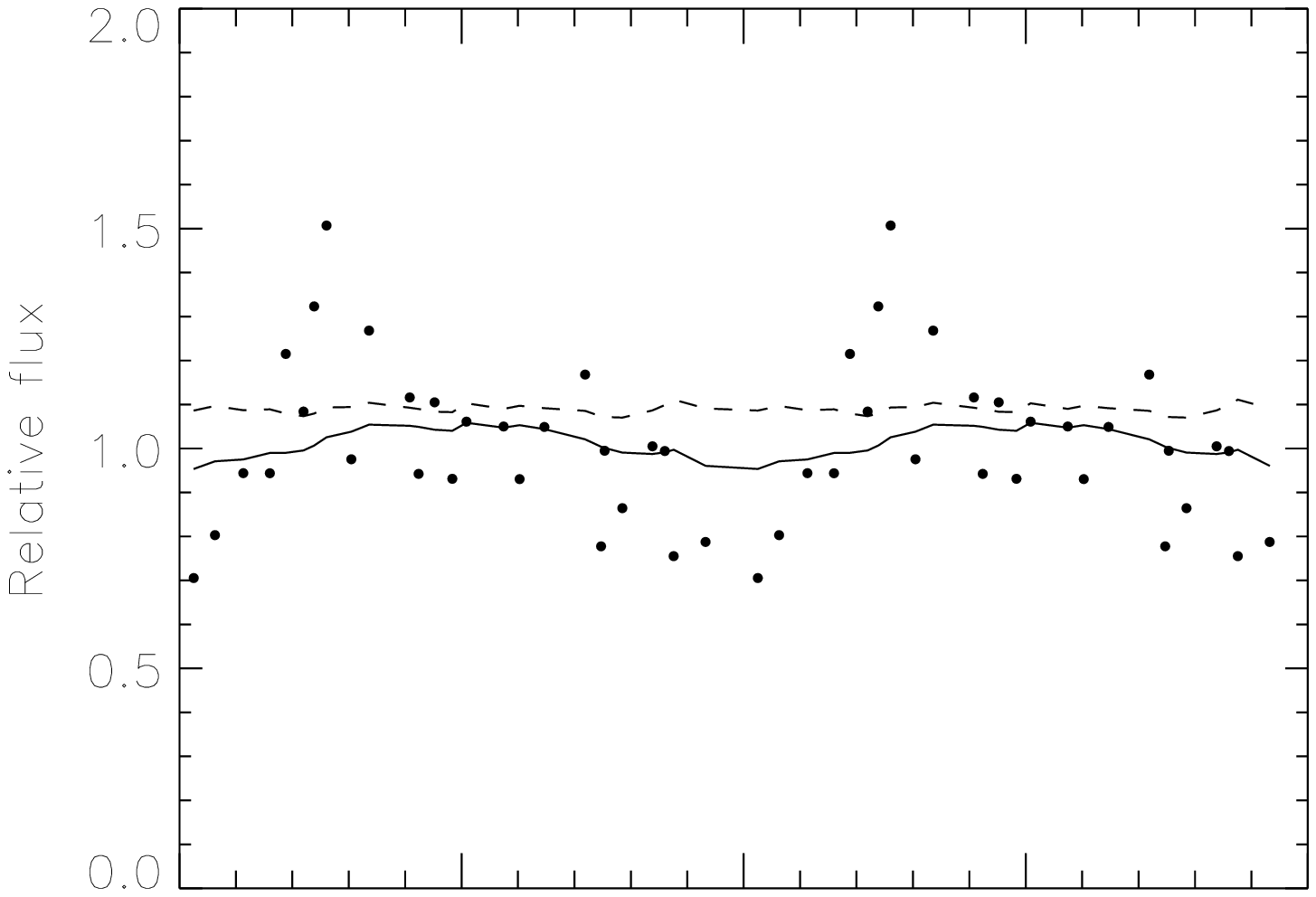}    
   \end{center}
  \end{minipage}
  \begin{minipage}{.45\hsize}
   \begin{center}
     \includegraphics[trim=0cm 1cm 0cm 1.0cm,width=.70\textwidth]{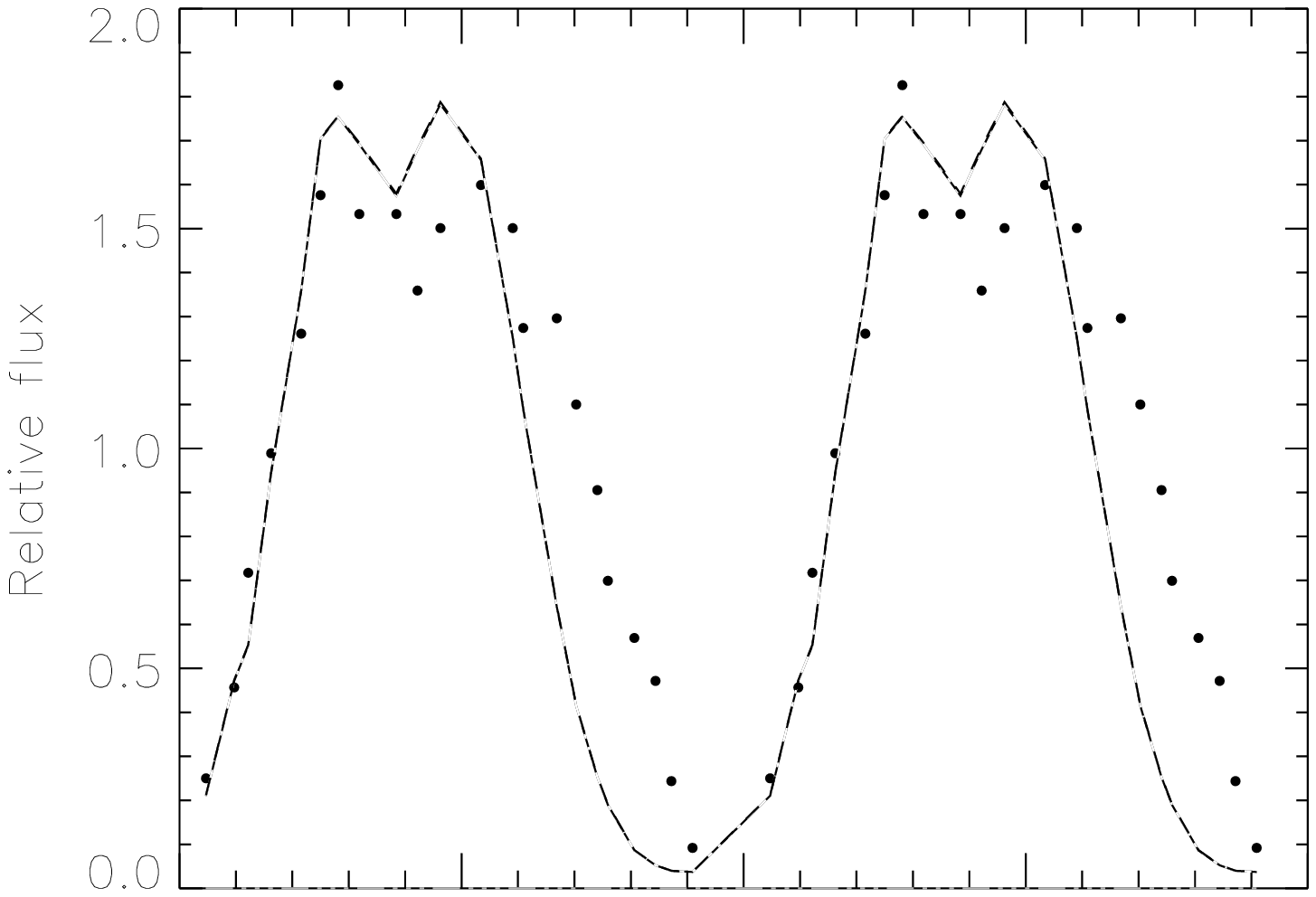}
     \includegraphics[trim=0cm 1cm 0cm 1cm,width=.70\textwidth]{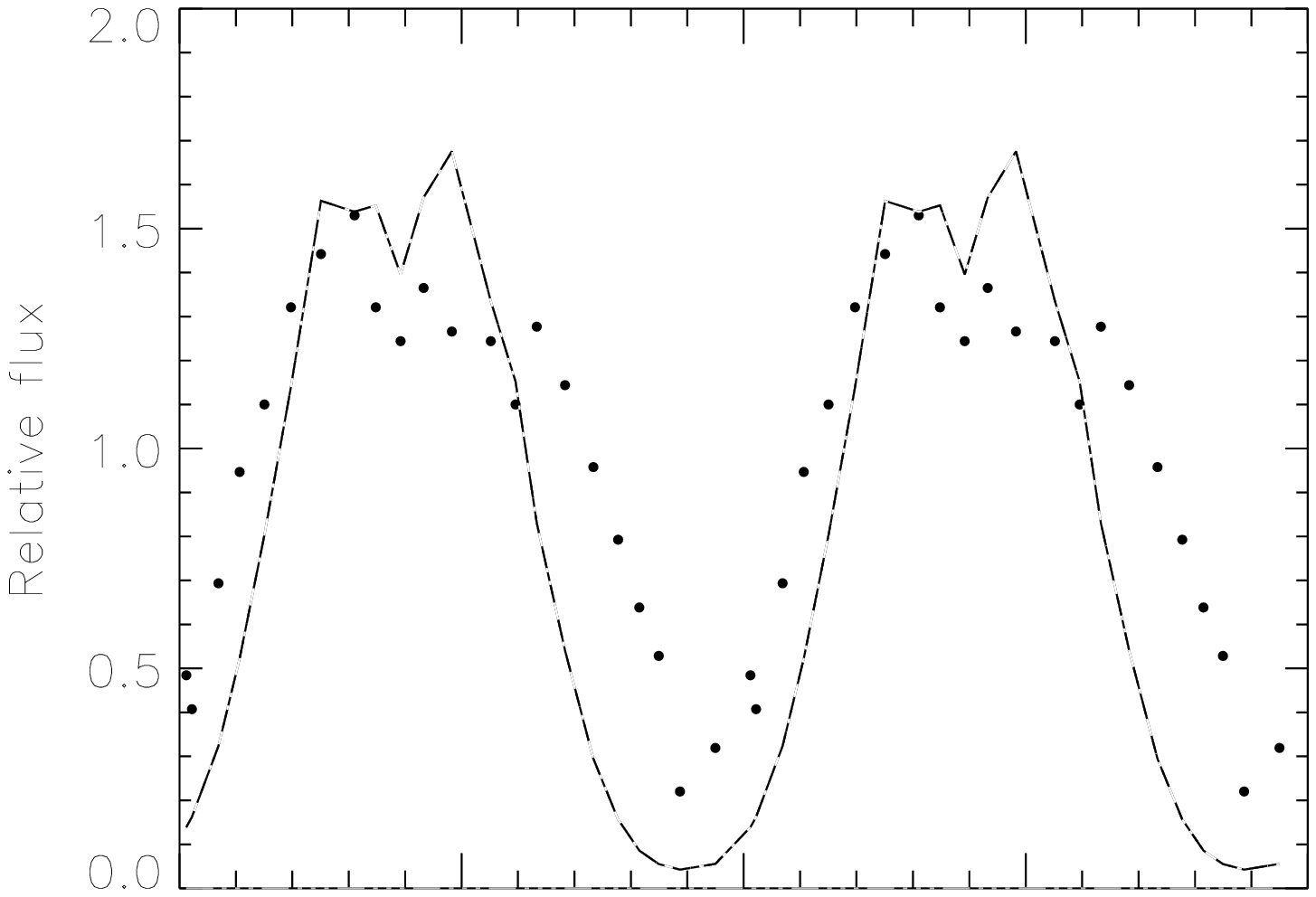}
     \includegraphics[trim=0cm 1cm 0cm 1cm,width=.70\textwidth]{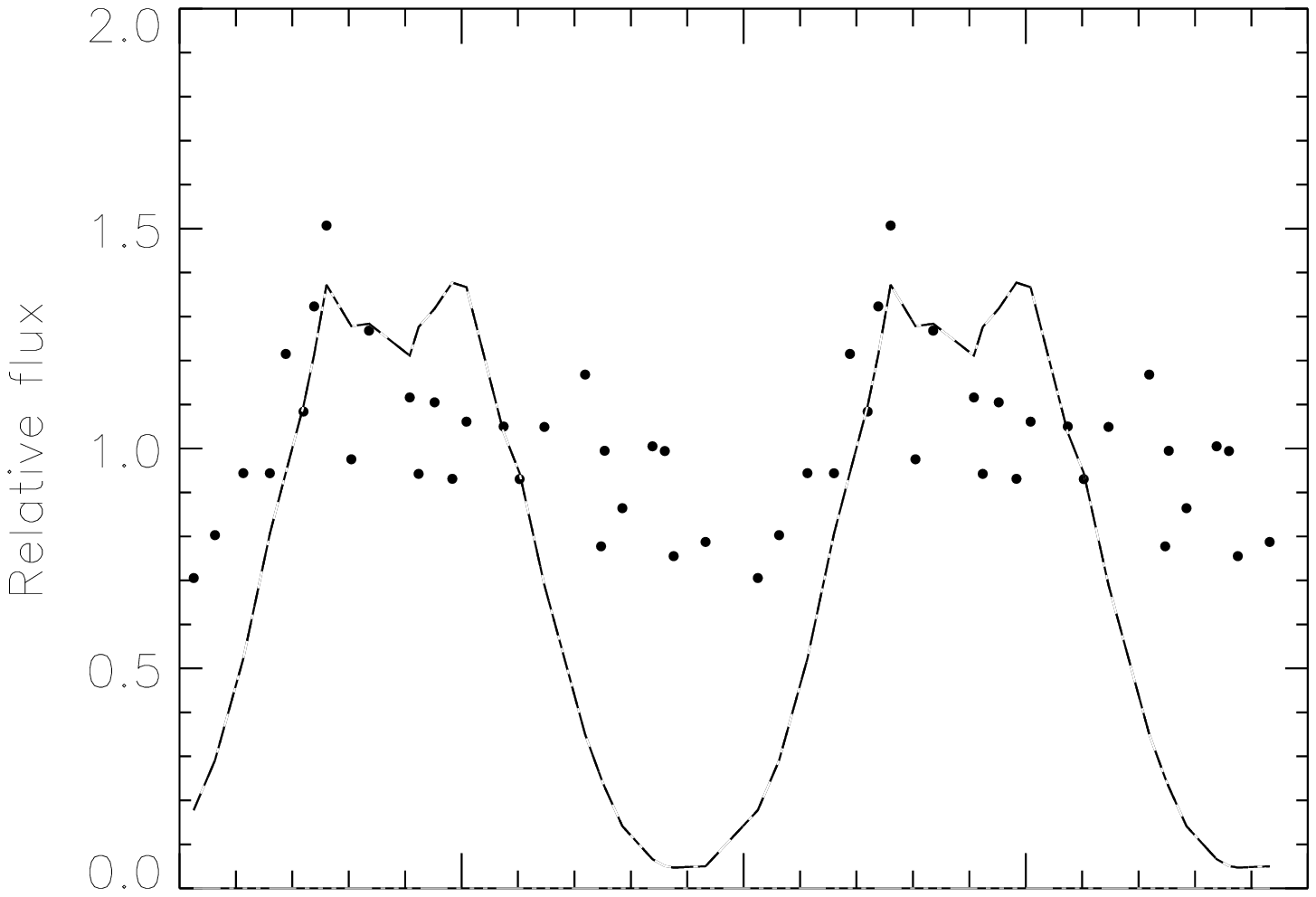} 
   \end{center}
  \end{minipage}
 \end{tabular}
\caption{Comparison between X-ray light curves of CP Tuc (points) and the modeled X-ray light curves. The abcissa is orbital phase and runs from 0 to 2. Left: Absorption scenario (solid line= model with absorption; dashed line= model without absorption). Right: Self-eclipse scenario.}
    \label{fig:lc}
\end{figure}

We have used {\sc cyclops-x} to obtain the X-ray light curves in each case using the parameters found by the optical fits. In doing that, we are considering that optical and X-ray data are representative of a same brightness state. Figure \ref{fig:lc} presents the observed \cite{mizaki:1996} and the modeled light curves in three bands: 0.7-2.3 keV (top), 2.3-6.0 keV (middle) and 6.0-10.0 keV (bottom).

\begin{figure}[ht]
   \begin{center}
 \includegraphics[clip=true,width=.80\textwidth]{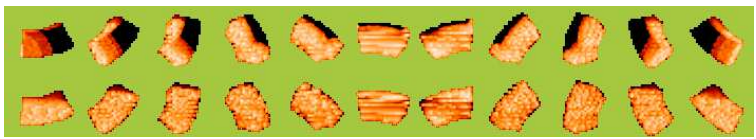}
       \caption{Images of the post-shock region in the absorption scenario. Up: considering absorption. Bottom: with no absorption. The images have different lookup tables.}
      \label{fig:geo1}
   \end{center}
\end{figure}

\textbf{Absorption scenario:} Figure \ref{fig:geo1} shows 11 images in different orbital phases produced by {\sc cyclops-x}. Specifically, it is shown the X-ray images of the post-shock region with (top) and without (bottom) the pre-shock absorption.  Figure \ref{fig:lc} shows that the absorption reproduces the energy dependence of the modulation, but the amplitude was insufficient to fit the data in the hardest bands. As the density of the pre-shock region has an upper limit defined by the minimum density of the post-shock region, an improvement of the X-ray fit may be achieved by an increase of the density in the post-shock region. However, it is necessary to check the influence on this on the optical modeling.

\begin{figure}[ht]
   \begin{center}
  \includegraphics[angle=90,clip=true,width=.80\textwidth]{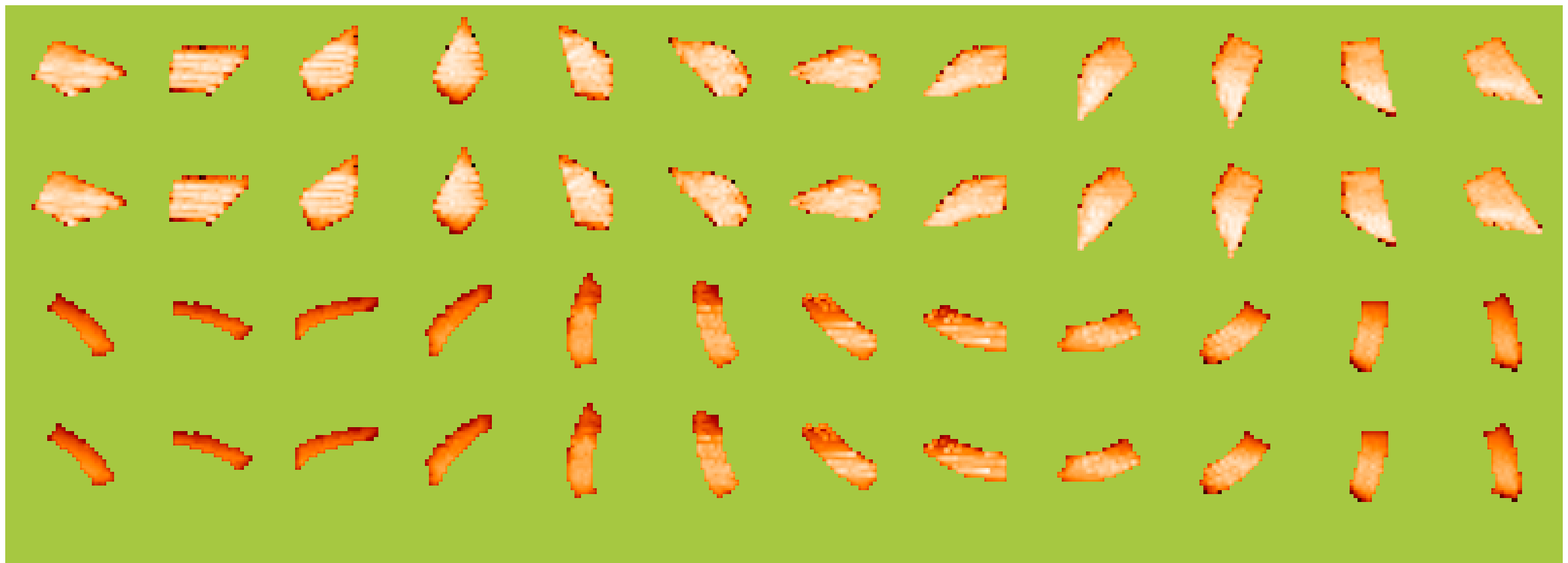}
       \caption{Images of the post-shock region in the self-eclipse scenario.}
      \label{fig:geo2}
   \end{center}
\end{figure}

\textbf{Self-eclipse scenario:} Figure \ref{fig:geo2} presents 11 images of the emitting region distributed along the orbital period. In this case, the absorption of the pre-shock region is not important, because it never is the line of sight. This model does not reproduce the dependence with energy as showed in Figure \ref{fig:lc}. How could this be improved? In this specific model, only the base of the accretion column is self-eclipsed. That region is responsible by most of the emission in soft energies, according to Equation \ref{emissividade} and the post-shock structure. A possibility to improve the X-ray fit is to change the temperature and the height of the post-shock region.  A higher column could present a larger range of temperatures and densities: it could provide a modulation more pronounced in the soft band than in the hard band. 

In both cases, the sets of parameters that produce good models in optical bands do not reproduce the X-ray data.  Nevertheless some important features of the data could be reproduced.
  
\section{Conclusions}

We present the {\sc cyclops-x}, a code that calculates the optical and X-ray emission of post-shock regions of polars. This code includes the absorption caused by the pre-shock region in a manner consistent with the geometry defined by the magnetic lines. We used the {\sc cyclops-x} to calculate the X-ray light curves of the two good optical models for CP Tuc. The optical emission of the two models are very similar, what is not true for the X-ray fluxes. This simple application shows the importance of consider as many data as possible to understand the post-shock region of a polar. We are planing to incorporate in {\sc cyclops-x} the simultaneous fit of X-ray and optical data and, then, apply this tool to the CP Tuc multiwavelength data. 

\section*{Acknowledgements}

This work was partially supported by Fapesp (CVR: Proc. 2010/06096-1; KMGS: Proc. 2008/09619-5).



\begin{thebibliography}{99}
\bibitem{mewe:1986} R. Mewe, J.R. Lemen and G. H.~J.~van der Oord, A\&A Supplement Series 65 (1986) 551.
\bibitem{arnoudi:1996} K.~A.~Arnaud, {\em ASP Conf. Ser. 101, Astronomical Data Analysis Software and Systems V}, (1996) Eds. G. Jacoby and J. Barnes, 17.
\bibitem{cropper:2000} M.~Cropper, K.~Wu and G.~Ramsay, New Astronomy Reviews 44 (2000) 57.
\bibitem{ramsay:2000} G.~Ramsay, S. B~.~Potter, M. Cropper, D.~A.~H.~Buckley and M.~K.~Harrop-Allin, MNRAS 316 (200) 225.
\bibitem{costa:2009} J.~E.~R.~Costa and C.~V.~Rodrigues, MNRAS 398 (2009) 240. 
\bibitem{gronenschild:1978} E.~H.~B.~M.~Gronenschild and R.~Mewe, A\&A  32 (1978) 283.
\bibitem{morrison:1983} R. Morrison and D. McCommon, Apj 270 (1983) 119.
\bibitem{mizaki:1996} K. Misaki, Y. Terashima, Y. Kamata, M. Ishida, H. Kunieda, and Y. Tawara, ApJ 470 (1996) L53. 
\bibitem{ramsay:1999} G. Ramsay, S.~B.~Potter, D.~A.~H.~Buckley, and P.~J.~Wheatley, MNRAS 306 (1999) 809. 
\end{thebibliography}
\end{document}